\documentclass[12 pt, draftclsnofoot, onecolumn]{IEEEtran}

\usepackage{graphicx}
\usepackage{amsmath}
\usepackage{amssymb}
\usepackage{graphics}
\usepackage{latexsym}
\usepackage[dvips]{epsfig}
\usepackage[nospace]{cite}
\usepackage{subfigure}
\usepackage{setspace}
\usepackage{color}
\usepackage{tabularx}

\newtheorem{Definition}{Definition}
\newtheorem{Lemma}{Lemma}
\newtheorem{Corollary}[Lemma]{Corollary}
\newtheorem{Proposition}[Lemma]{Proposition}
\newtheorem{Theorem}{Theorem}

\newtheorem{Remark}{Remark}

\usepackage[ colorlinks = true,
             linkcolor = blue,
             urlcolor  = blue,
             citecolor = red,
             anchorcolor = green,
]{hyperref}

\allowdisplaybreaks[1]

\def\Pr{{\mathrm{Pr}}}
\def\E{{\mathrm E}}
\def\Var{{\mathrm {Var}}}

\begin{document}
%
\title{\huge Non-Asymptotic Achievable Rates for Energy-Harvesting Channels using Save-and-Transmit}
%
%
%


\author{Silas~L.~Fong$^\dagger$, Vincent~Y.~F.~Tan$^\dagger$ and Jing Yang$^*$
\thanks{$^\dagger$Silas~L.~Fong and Vincent~Y.~F.~Tan are with the Department of Electrical and Computer Engineering, National University of Singapore (NUS), Singapore (e-mail: \texttt{\{silas\_fong,vtan\}@nus.edu.sg}). Vincent~Y.~F.~Tan  is also with the Department of Mathematics, NUS. }
\thanks{$^*$Jing Yang is with the Department of Electrical Engineering at the University of Arkansas, AR, USA (email: \texttt{jingyang@uark.edu})}}

\maketitle

\begin{abstract}
This paper investigates the information-theoretic limits of energy-harvesting (EH) channels in the finite blocklength regime. The EH process is characterized by a sequence of i.i.d.\ random variables with finite variances. We use the save-and-transmit strategy proposed by Ozel and Ulukus (2012) together with Shannon's non-asymptotic achievability bound to obtain lower bounds on the achievable rates for both additive white Gaussian noise channels and discrete memoryless channels under EH constraints. The first-order terms of the lower bounds of the achievable rates are equal to $C$ and the second-order (backoff from capacity) terms are proportional to $-\sqrt{ \frac{\log n}{n} }$, where $n$ denotes the blocklength and $C$ denotes the capacity of the EH channel, which is the same as the capacity without the EH constraints. The constant of proportionality of the backoff term is found and qualitative interpretations are provided. 
\end{abstract}


\IEEEpeerreviewmaketitle

\flushbottom

\section{Introduction} \label{Introduction}

The energy-harvesting (EH) channel consists of one source equipped with an energy buffer, and one destination.  For simplicity, in this paper, we assume that the buffer has infinite capacity. At each discrete time $k \in \{1,2,\ldots\}$, a random amount of energy $E_k\in [0, \infty)$ arrives at the buffer and the source transmits a symbol $X_k\in (-\infty, \infty)$ such that
 \begin{equation}
 \sum_{\ell=1}^k X_\ell^2 \le \sum_{\ell=1}^k E_\ell   \qquad\mbox{almost surely}. \label{EHResultIntro}
 \end{equation}
 This implies that the total harvested energy $\sum_{\ell=1}^k E_\ell$ must be no smaller than the energy of the codeword $\sum_{\ell=1}^k X_\ell^2$  at every discrete time $k$ for transmission to take place successfully.
 We assume that $\{E_\ell\}_{\ell=1}^\infty$ are independent and identically distributed (i.i.d.) non-negative random variables, where $\E[E_1]=P$ and $\E[E_1^2]<+\infty$. The destination receives
\begin{equation}
 Y_k = X_k + Z_k
 \end{equation}
at time slot~$k$ for each $k\in\mathbb{N}$ where $\{Z_k\}_{k=1}^{\infty}$ are i.i.d.\ standard normal random variables. We refer to  the above EH channel as the \textit{additive white Gaussian noise (AWGN) EH channel}. It was shown by Ozel and Ulukus~\cite{ozel12} that the capacity of the AWGN EH channel is
 \begin{equation}
 C=\frac{1}{2}\log (1+P) \qquad\mbox{nats per channel use},   \label{eqn:capacity_awgn}
 \end{equation}
 where $P=\E[E_1]$ is the expectation of the harvested energy for each energy arrival.
The AWGN EH channel models real-world, practical situations where energy may not be fully available at the time of transmission and its unavailability may result in the transmitter not being able to put out the desired codeword. This model is applicable in large-scale sensor networks where  each node is equipped with an EH device that collects a stochastic amount of energy. See~\cite{ulukus15} for a comprehensive review of recent advances in EH wireless communications.

Observe that  the capacity in \eqref{eqn:capacity_awgn} is unchanged vis-\`a-vis the AWGN channel {\em without} the EH constraints. The capacity is an {\em asymptotic} concept, defined as the number of channel uses $n$ is allowed to tend to infinity. The result of Ozel and Ulukus~\cite{ozel12} thus masks the higher-order rate-losses that may arise due to the EH constraints. However, in many real-world applications, delay or latency constraints are present, e.g., we cannot wait infinitely long to decode the transmitted codeword.  This begs the natural question: {\em Information-theoretically, and from a finite blocklength (non-asymptotic) \cite{PPV10, Tan_FnT}  perspective, what are the rate-losses for various channels when we take the EH constraints into account?} This is what this paper investigates both for the AWGN EH channel and its discrete memoryless counterpart. One of the main takeaways from this work is that, compared to when EH constraints are not present (cf.~\cite{PPV10}), there can potentially be a significant  backoff from capacity at moderate blocklengths if  one uses the save-and-transmit  strategy~\cite{ozel12} to take the EH constraints in account. 

 \subsection{Main Contribution}
The contributions of this paper are threefold:

\begin{enumerate}
\item First, we prove achievable finite blocklength bounds for  EH channels under the constraint in \eqref{EHResultIntro} based on the save-and-transmit strategy of~\cite{ozel12}. During the saving phase of the save-and-transmit strategy, we save energy for a certain number  of time slots. During this period, no information is transmitted. Subsequently, during the transmission phase,  we use the remaining time slots to send information. By carefully developing various concentration bounds to control the probability that the available energy is insufficient to support the transmitted codeword during the transmission phase (i.e., that $\sum_{\ell=1}^k E_\ell < \sum_{\ell=1}^k X_\ell^2$), we show that the backoff from capacity $C$ at a blocklength $n$ is no larger than $O(\sqrt{n^{-1}\log n} )$. In other words, the maximum number of codewords  we can transmit over $n$ channel uses with average probability of error no larger than $\varepsilon$, denoted by $M_{n,\varepsilon}^*$, satisfies
\begin{equation}
\frac{1}{n}\log M_{n,\varepsilon}^* \ge C - \kappa\sqrt{\frac{\log n }{n}}.
\end{equation}
 We also identify the implied constant $\kappa$ and provide qualitative interpretations.  Such an analysis for noisy channels was not available prior to the present work. Furthermore, by scrutinizing the analysis of Ozel and Ulukus~\cite{ozel12} for AWGN channels, one can also deduce that the backoff from capacity is no larger than $O(n^{-1/2} \log n )$. Thus, our analysis results in a slightly smaller (tighter) backoff than what was implied by the authors in~\cite{ozel12}.

\item Second, our analysis only requires minimal statistical assumptions on the EH process $\{E_\ell\}_{\ell=1}^\infty$. Indeed, apart from assuming that the process is i.i.d., we only assume that the second moment of the EH  random variable $E_\ell$ is bounded, i.e.,
\begin{equation}
\E [ E_\ell^2  ]<\infty,\qquad \forall \, \ell\in\mathbb{N}.
\end{equation}
In previous results such as \cite[Lemmas 1 \& 2]{ozel12}, more restrictive assumptions on $E_\ell$ were made, e.g., that $\E \big[e^{E_\ell^\gamma} \big]$ is bounded for some $\gamma \in (0,1)$. This assumption may be hard to verify in practice.

\item Finally, we study both AWGN and     discrete memoryless EH  channels. Previous work on finite blocklength analysis for EH channels was performed by Yang~\cite{yangEH2014} for {\em noiseless binary} channels. The analysis required the use of sophisticated martingale convergence theorems and renewal process theory. Our analysis is comparatively simple, making use of basic probability techniques such as Markov's and Chebyshev's inequalities. Furthermore, we also analyze finite blocklength fundamental limits of {\em noisy} channels.
\end{enumerate}

\subsection{Related Work}
Information-theoretic characterizations of EH communication channels have been investigated recently. As energy arrives randomly to the transmitter, codewords must satisfy the cumulative stochastic energy constraints. The impact of the stochastic energy supply on the channel capacity was characterized for an additive white Gaussian noise (AWGN) channel with an i.i.d.\ EH process in~\cite{ozel12} and with a stationary ergodic EH process in~\cite{RSV14}. The aforementioned studies showed that with an unlimited battery, the capacity of the AWGN channel with stochastic energy constraints is equal to the capacity of the same channel under an average power constraint, as long as the average power equals the average recharge rate of the battery. 

Using Shannon's coding scheme for channels with causal state information at the encoder~\cite{Shannon:1958:CSI}, the zero battery case for the same problem was discussed in \cite{Ozel:2011:Asilomar}. Jog and Anatharam~\cite{Jog:ISIT:2014}  characterized the capacity of the AWGN channel with a finite battery when energy arrivals are deterministic. Dong, Farnia and \"{O}zg\"{u}r~\cite{Ozgur:NOE:2015} provided an approximation to the capacity with bounded guarantee on the approximation gap for i.i.d.\ Bernoulli energy arrivals.  In recent work, Shaviv, Nguyen and \"{O}zg\"{u}r~\cite{SNOzgur:ISIT:2015} provided an $n$-letter expression for the channel capacity with and without causal and noncausal energy arrival information at the transmitter and/or the receiver. In addition, Shaviv and \"{O}zg\"{u}r~\cite{SOzgur:ISIT:2015} investigated a similar problem with a Bernoulli recharge process.

Mao and Hassibi~\cite{Mao:2013} investigated the capacity of an energy-harvesting transmitter with finite battery over a discrete memoryless channel (DMC). It was shown that the capacity can be described using the Verd\'u-Han general framework~\cite{Verdu:2006:GFC}. If the transmitted symbol only depends on the currently available energy, the system reduced to a finite-state channel. However, it was analytically intractable to explicitly characterize the capacity, and even the lower bound of the capacity can only be evaluated numerically.
A special scenario of the same problem, namely the capacity of noiseless binary channel with binary energy arrivals and unit-capacity battery, was discussed in~\cite{TOYU:2013:ISIT}. The channel was shown to be equivalent to an additive geometric-noise timing channel with causal information of the noise available at the transmitter. Achievable strategies were proposed along with upper bounds, which were then improved in \cite{TutuncuogluOYU14}. Ozel {\em et al.} \cite{Ozel:ISIT:2015} considered a noiseless binary energy harvesting channel with on-off fading.
%

As mentioned above, finite blocklength analysis for EH channels was only considered previously by Yang~\cite{yangEH2014}. However, the channel considered therein is noiseless and has binary inputs and binary outputs. Our framework is considerably more general and we consider noisy discrete as well as Gaussian channels from a finite blocklength perspective. The study of finite blocklength fundamental limits in Shannon-theoretic problems was undertaken by Polyanskiy, Poor and Verd\'u~\cite{PPV10}. Such a study is useful as it provides guidelines regarding the  required backoff from the asymptotic fundamental limit (capacity) when one operates at finite blocklengths. For a survey, please see~\cite{Tan_FnT}.
\subsection{Paper Outline}
This paper is organized as follows. The notation used in this paper is described in the next subsection. Section~\ref{sectionDefinition} states the formulation of the AWGN EH channel and presents our main theorem. Numerical results are also provided. Section~\ref{sectionSaveAndTransmit} describes the save-and-transmit strategy and proves our main theorem. More specifically, we use Shannon's achievability bound~\cite{sha57} to prove an achievable rate for the save-and-transmit strategy. Section~\ref{sectionFiniteAlphabet} provides and proves an analogous result for  discrete memoryless EH channels. Concluding remarks are provided in Section~\ref{sec:conclusion}.

 \subsection{Notation} \label{sectionNotation}
We let $\boldsymbol{1}(\mathcal{E})$ be the indicator  function of the set  $\mathcal{E}$.
We use the upper case letter~$X$ to denote an arbitrary (discrete or continuous) random variable with alphabet $\mathcal{X}$, and use a lower case letter $x$ to denote a realization of~$X$.
We use $X^n$ to denote the random tuple $(X_1, X_2, \ldots, X_n)$. 

The following notations are used for any arbitrary random variables~$X$ and~$Y$ and any real-valued function $g$ with domain $\mathcal{X}$. We let $p_{X,Y}$ and $p_{Y|X}$ denote the probability distribution of $(X,Y)$
 and the conditional probability distribution of $Y$ given $X$ respectively.
We let $p_{X,Y}(x,y)$ and $p_{Y|X}(y|x)$ be the evaluations of $p_{X,Y}$ and $p_{Y|X}$ respectively at $(X,Y)=(x,y)$. 
To make the dependence on the distribution explicit, we let $\Pr_{p_X}\{ g(X)\in\mathcal{A}\}$ denote $\int_{x\in \mathcal{X}} p_X(x)\mathbf{1}\{g(x)\in\mathcal{A}\}\, \mathrm{d}x$ for any set $\mathcal{A}\subseteq \mathbb{R}$.
The expectation and the variance of~$g(X)$ are denoted as
$
\E_{p_X}[g(X)]$ and
$
 \Var_{p_X}[g(X)]=\E_{p_X}[(g(X)-\E_{p_X}[g(X)])^2]$
 respectively. 
 We let $\mathcal{N}(\,\cdot\, ;\mu,\sigma^2): \mathbb{R}\rightarrow [0,\infty)$ denote the probability density function of a Gaussian random variable whose mean and variance are $\mu$ and $\sigma^2$ respectively, i.e., \begin{equation}
\mathcal{N}(z;\mu,\sigma^2)\triangleq\frac{1}{\sqrt{2\pi \sigma^2}}\exp\bigg(-\frac{(z-\mu)^2}{2\sigma^2} \bigg). \label{eqnNormalDist}
\end{equation}
We will take all logarithms to base $e$ throughout this paper. 

\section{Additive White Gaussian Noise Energy-Harvesting Channel}
\label{sectionDefinition}
\subsection{AWGN EH Model}
The AWGN EH  channel consists of one source and one destination, denoted by $\mathrm{s}$ and $\mathrm{d}$ respectively. Node~$\mathrm{s}$ transmits information to node~$\mathrm{d}$ in $n$ time slots as follows. Node~$\mathrm{s}$ chooses message
$
W
$
 and sends $W$ to node~$\mathrm{d}$, where $W$ is uniformly distributed over $\{1, 2, \ldots, M\}$ and $M=|\mathcal{W}|$. Then for each $k\in \{1, 2, \ldots, n\}$, node~$\mathrm{s}$ transmits $X_{k}\in \mathbb{R}$ and node~$\mathrm{d}$ receives $Y_k\in \mathbb{R}$ in time slot~$k$. Let $E_1, E_2, \ldots, E_n$ be i.i.d.\ random variables that satisfy
$
\Pr\{E_1< 0\}=0$,
$
\E[E_1]=P$
and
$
\E[E_1^2]<\infty$.
We assume the following for each $k\in\{1, 2, \ldots, n\}$:
\begin{enumerate}
\item[(i)] $E_k$ and $(W, E^{k-1}, X^{k-1}, Y^{k-1})$ are independent, i.e.,
\begin{align}
p_{W, E^{k}, X^{k-1}, Y^{k-1}} = p_{E_{k}}p_{W, E^{k-1}, X^{k-1}, Y^{k-1}} \label{assumption(i)}
\end{align}
    \item[(ii)] Every codeword $X^n$ transmitted by~$\mathrm{s}$ should satisfy
\begin{equation}
\Pr\left\{\sum_{\ell=1}^k X_\ell^2 \le \sum_{\ell=1}^k E_\ell\right\}=1 \label{eqn:eh}
\end{equation}
for each $k\in\{1, 2, \ldots, n\}$.
  \end{enumerate}
   After~$n$ time slots, node~$\mathrm{d}$ declares~$\hat W$ to be the transmitted~$W$ based on $Y^n$. Formally, we define a code as follows:
\begin{Definition} \label{defCode}
An {\em $(n, M)$-code} consists of the
following:
\begin{enumerate}
\item A message set
$
\mathcal{W}\triangleq \{1, 2, \ldots, M\}
$
 at node~$\mathrm{s}$. Message $W$ is uniform on $\mathcal{W}$.

\item A sequence of  encoding functions
$
f_k : \mathcal{W}\times \mathbb{R}_+^{k}\rightarrow \mathbb{R}
$
 for each $k\in\{1, 2, \ldots, n\}$, where $f_k$ is the encoding function for node~$\mathrm{s}$ at time slot $k$  for encoding $X_k$ such that
$
X_k=f_k (W, E^{k})
$
and \eqref{eqn:eh} holds.
\item A decoding function
$
\varphi :
\mathbb{R}^{n} \rightarrow \mathcal{W},
$
for decoding $W$ at node~$\mathrm{d}$ by producing
$
 \hat W = \varphi(Y^{n})$.
\end{enumerate}
\end{Definition}
\begin{Definition}\label{defAWGNchannel}
The {\em AWGN EH channel} is characterized by $q_{Y|X}$ such that the following holds for any $(n, M)$-code: For each $k\in\{1, 2, \ldots, n\}$,
\begin{align}
p_{W, E^k, X^k, Y^k}
 = p_{W, E^k, X^k, Y^{k-1}}p_{Y_k|X_k} \label{memorylessStatement*}
\end{align}
where
\begin{equation}
p_{Y_k|X_k}(y_k|x_k) = q_{Y|X}(y_k|x_k) = \mathcal{N}(y_k-x_k; 0,1) \label{defChannelInDefinition*}
\end{equation}
for all $x_k$ and $y_k$.
Since $p_{Y_k|X_k}$ does not depend on~$k$ by \eqref{defChannelInDefinition*}, the channel is stationary.
\end{Definition}

 For any $(n, M)$-code defined on the AWGN EH channel, let $p_{W,E^n, X^n, Y^n, \hat W}$ be the joint distribution induced by the code. We can factorize $p_{W,E^n, X^n, Y^n, \hat W}$ as follows:
\begin{align}
 p_{W,E^n, X^n, Y^n, \hat W}
&\stackrel{\text{(a)}}{=} p_{W,E^n, X^n, Y^n}p_{\hat W |Y^n}  \notag\\
&= p_{W} \left(\prod_{k=1}^n  p_{E_k|W , E^{k-1} ,X^{k-1} , Y^{k-1}}p_{X_k,Y_k|W, E^k ,X^{k-1} , Y^{k-1}}\right)  \times p_{\hat W |Y^n} \notag\\
&\stackrel{\eqref{assumption(i)}}{=} p_{W} \left(\prod_{k=1}^n p_{E_k}p_{X_k,Y_k|W, E^k,X^{k-1}, Y^{k-1}}\right)p_{\hat W |Y^n} \notag\\
&= p_W\!\! \left(\prod_{k=1}^n \!  p_{E_k} p_{X_k|W, E^k\!, X^{k-1}\!, Y^{k-1}}p_{Y_k|W, E^k\!, X^k\!, Y^{k-1}} \!\!\right)\!p_{\hat W |Y^n} \notag\\
& \stackrel{\text{(b)}}{=} p_W \left(\prod_{k=1}^n p_{E_k}p_{X_k|W, E^k} p_{Y_k|W, E^k, X^k, Y^{k-1}} \right)p_{\hat W |Y^n}\notag\\
& \stackrel{\eqref{memorylessStatement*}}{=} p_W \left(\prod_{k=1}^n p_{E_k} p_{X_k|W, E^k} p_{Y_k|X_k}\right)p_{\hat W |Y^n}. \label{memorylessStatement}
\end{align}
where
\begin{enumerate}
\item[(a)] uses the fact by Definition~\ref{defCode} that $\hat W$ is a function of $Y^n$.
\item[(b)] uses the fact by Definition~\ref{defCode} that $X_k$ is a function of $(W, E^k)$ for each $k\in\{1, 2, \ldots, n\}$.
\end{enumerate}
\begin{Definition} \label{defErrorProbability}
For an $(n, M)$-code defined on the AWGN EH channel, we can calculate, according to \eqref{memorylessStatement}, the \textit{average probability of decoding error} defined as $\Pr\big\{\hat W \ne W\big\}$.
We call an $(n, M)$-code with average probability of decoding error no larger than $\varepsilon$ an {\em $(n, M, \varepsilon)$-code}.
\end{Definition}
\begin{Definition} \label{defAchievableRate}
Let $\varepsilon\in [0,1)$ be a real number. A rate $R$ is \textit{$\varepsilon$-achievable} for the AWGN EH channel if there exists a sequence of $(n, M_n, \varepsilon_n)$-codes such that
\begin{equation}
\liminf_{n\rightarrow \infty}\frac{1}{n}\log M_n \ge R \quad\mbox{and}\quad \limsup_{n\rightarrow \infty}\varepsilon_n \le \varepsilon.
\end{equation}
\end{Definition}

\begin{Definition}\label{defCapacity}
Let $\varepsilon\in [0,1)$ be a real number. The {\em $\varepsilon$-capacity} for the AWGN EH channel, denoted by $C_\varepsilon$, is defined to be
$
C_\varepsilon \triangleq \sup\{R: R\text{ is $\varepsilon$-achievable}\}$.
\end{Definition}
%
\subsection{Main Result}
The following theorem is the main result in this paper. The proof is contained in Section~\ref{sectionSaveAndTransmit} after we illustrate the result  numerically  in Section~\ref{sec:numerical}.
\begin{Theorem}\label{thmMainResultLowerBound}
Let $\varepsilon\in(0,1)$, and define
\begin{equation}
a\triangleq \max\left\{\E_{p_{E_1}}[E_1^2], 12\sqrt{2} P^2\right\}. \label{defAInTheorem}
\end{equation}
Suppose~$n\ge 3$ is a sufficiently large integer such that
 \begin{equation}
 \frac{n}{\log n} \ge \max \left\{\frac{\E_{p_{E_1}}[E_1^2]}{P^2}, 12\sqrt{2}\right\}, \label{sufficientlyLargeNinProof}
 \end{equation}
 \begin{equation}
 n\ge \left(\log\left(\frac{2+\varepsilon}{\varepsilon^2}\right)\right)^4 \label{sufficientlyLargeNinProof*}
  \end{equation}
  and
   \begin{equation}
 n\log n \ge \frac{e^{0.4}(2+\varepsilon)}{\varepsilon}.\label{sufficientlyLargeNinProof**}
  \end{equation}
Then, there exists an~$(n + m, M, \varepsilon)$-code such that
\begin{align}
&\log M \ge  \frac{n}{2}\log (1+P)- \sqrt{\frac{(2+\varepsilon)n P}{\varepsilon(P+1)}} - n^{\frac{1}{4}}-1 \label{thmMainResultStatement**}
\end{align}
where $m\triangleq \left\lceil \frac{6\sqrt{a n\log n}}{P} \right\rceil$ denotes the length of the initial saving period before any transmission occurs and~$n$ denotes the length of the actual transmission period.
In particular, there exists an~$(n^*, M, \varepsilon)$-code with $n^* \triangleq n + m$ such that
\begin{align}
\log M &\ge \frac{n^*}{2}\log (1+P)- \frac{3\log (1+P)\sqrt{a n^*\log n^*}}{P}  \notag\\
 &\quad - \sqrt{\frac{(2+\varepsilon)n^* P}{\varepsilon(P+1)}} - (n^*)^{\frac{1}{4}}-\frac{1}{2}\log(1+P)-1\,. \label{thmMainResultStatement}
\end{align}
\end{Theorem}
\begin{Remark}
Since $\frac{1}{n^*}\sum_{k=1}^{n^*} E_k$ converges to $\E_{p_{E_1}}\left[E_1\right]=P$ with probability one by the strong law of large numbers, it follows from the power constraint \eqref{eqn:eh} and the strong converse theorem for the AWGN channel~\cite{Yoshihara, kost15} that the $\varepsilon$-capacity of the AWGN EH channel is upper bounded by $\frac{1}{2}\log (1+P)$. Therefore, by normalizing both side of \eqref{thmMainResultStatement} by $n$ and taking the limit, we see that  Theorem~\ref{thmMainResultLowerBound} implies that  the $\varepsilon$-capacity is
\begin{equation}
C_\varepsilon=\frac{1}{2}\log (1+P),\qquad\forall\, \varepsilon\in [0,1).
\end{equation}
\end{Remark}

\begin{Remark}
The investigation of the save-and-transmit scheme by Ozel and Ulukus in \cite[Lemma 2]{ozel12} implies that $\frac{n^*}{2}\log (1+P)-O(\sqrt{n^*}(\log n^*)^\alpha)$ nats is  achievable over $n$ channel uses for any $\alpha>1$ and for $n\to\infty$. Theorem~\ref{thmMainResultLowerBound} improves the lower bound of the second-order term because the backoff term improves from  $-\sqrt{n^*}(\log n^*)^\alpha$ to $-\sqrt{n^*\log n^*}\, $.
\end{Remark}
\begin{Remark} \label{remark3}
It follows from \eqref{defAInTheorem} and \eqref{thmMainResultStatement} that the coefficient of the second-order term achieved by the save-and-transmit strategy is at least
\begin{equation}
\nu \triangleq  \frac{-3\log (1+P)\sqrt{\max\left\{\E_{p_{E_1}}[E_1^2], 12\sqrt{2} P^2\right\}}}{P}.
\end{equation}
Note that \eqref{thmMainResultStatement} is a direct consequence of \eqref{thmMainResultStatement**} and the derivation can be found in the equations from~\eqref{eqn11InCalculationErrorProb**} to~\eqref{eqn12InCalculationErrorProb} in the proof of the theorem. By inspecting the aforementioned derivation, we see that the second-order term in~\eqref{thmMainResultStatement} is due to the saving period only, which means the second-order term~$\nu$ is affected by the length of the saving period~$m$ alone (but not~$\varepsilon$).
  As $P$ increases, the magnitude of $\nu$ increases and hence a longer saving period is required to guarantee a certain probability of {\em outage}, namely that the transmitted codeword does not satisfy all the EH constraints. This corroborates the fact that as~$P$ increases, the variance of each Gaussian codeword increases and hence a longer saving period is required to maintain a certain outage probability. Similarly, as $\E_{p_{E_1}}[E_1^2]$ increases while~$P$ is fixed, the variance of the energy arrival process is larger and hence a longer saving period is required to maintain a certain outage probability.
\end{Remark}

\begin{Remark} \label{remark4}
We use Chebyshev's inequality to obtain the third term in the asymptotic expansion in \eqref{thmMainResultStatement}, i.e., the one proportional to  $\sqrt{n^*}$. One could also use the Berry-Esseen central limit theorem~\cite[Ch.\ XVI.5]{feller} to obtain a possibly better bound. However, more terms, such as the third absolute moments of certain random variables, would be involved. In addition, only the coefficient of the third-order term in~\eqref{thmMainResultStatement} can be improved slightly by using Berry-Esseen's theorem instead of Chebshev's inequality, and the improvement is minimal compared with the first- and second-order terms in~\eqref{thmMainResultStatement}. Thus, we have chosen to present a simpler achievability bound.
\end{Remark}

\subsection{Numerical Results} \label{sec:numerical}
\begin{figure*}[t]
 \centering
   \subfigure[{$\E[E_1]=P=3 \,\text{dB}$ and $\Var[E_1]=100$}]{\includegraphics[width=.475\columnwidth]{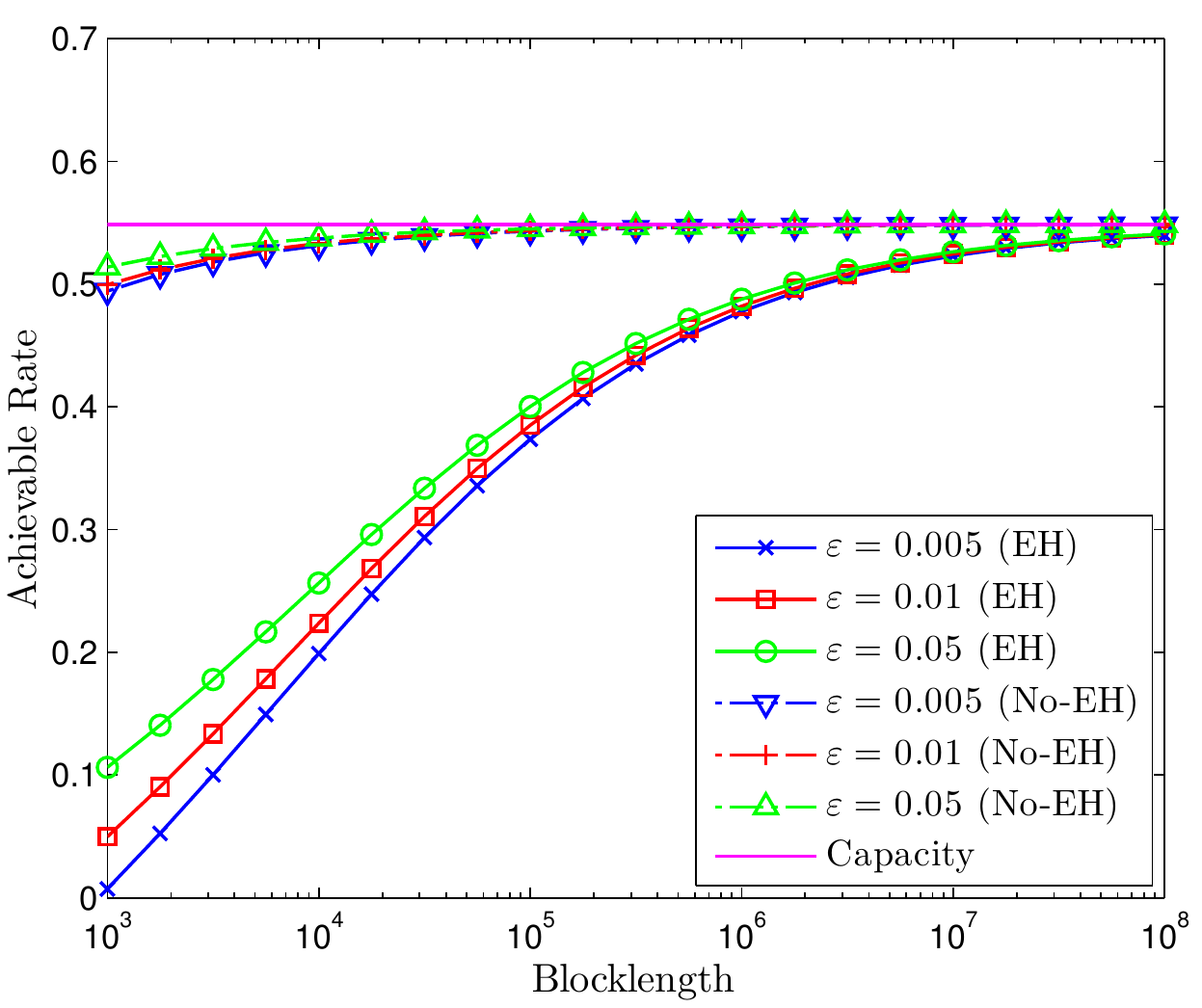}}
  \subfigure[{$\E[E_1]=P=3 \,\text{dB}$ and $\varepsilon = 0.01$}]{\includegraphics[width=.475\columnwidth]{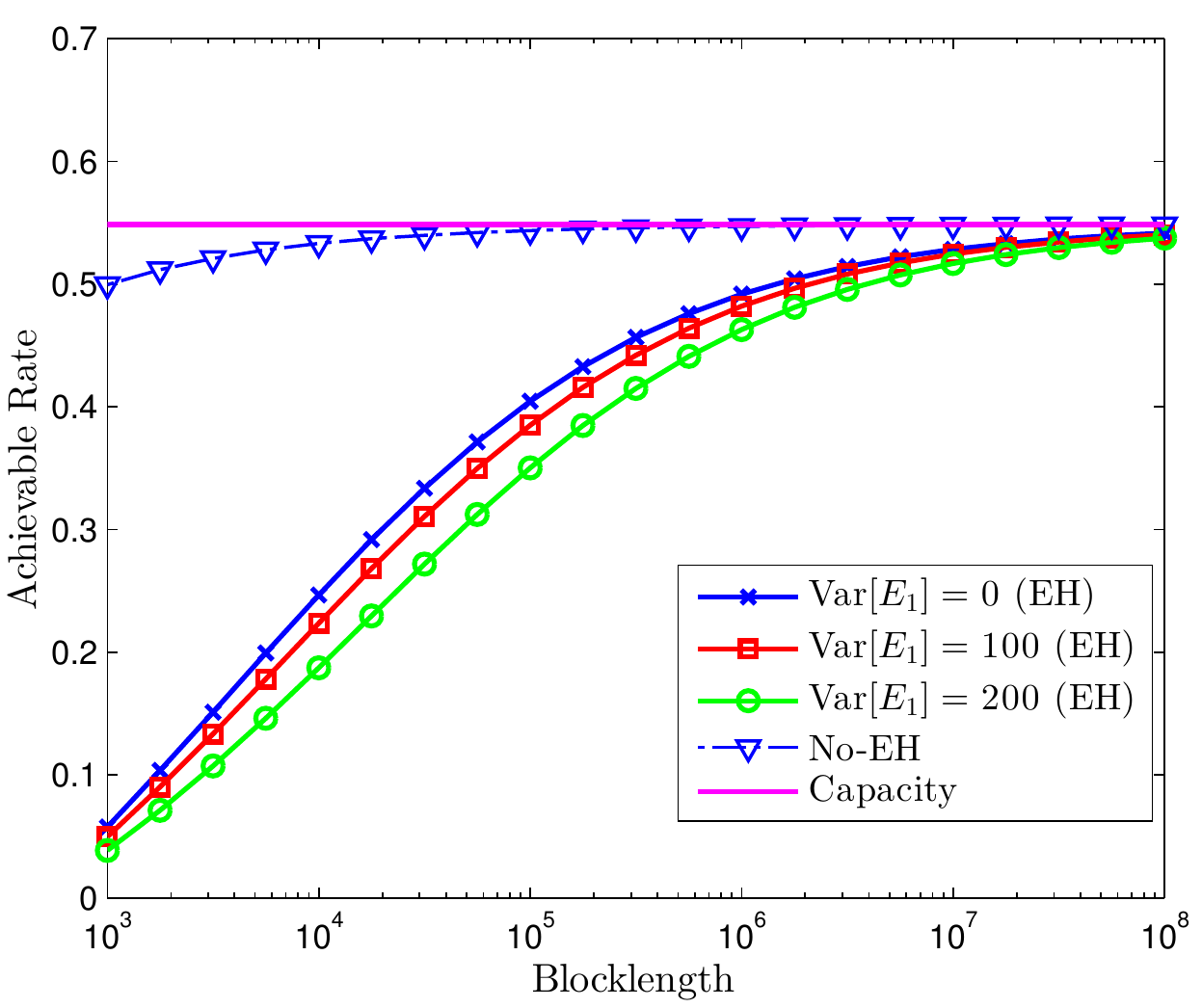}}
\caption{Achievable rates  for the save-and-transmit scheme in~\eqref{rValueNumerical}  when the error probability is varied (left) and when the variance of the EH process is varied (right). On the plot on the right (Figure 1(b)) for the No-EH line, the peak power (cf.\ \eqref{peakPowerConstraint}) is kept at~$P=3\,\text{dB}$}  \label{figure1}
\vspace{-0.3 in}
\end{figure*}
In this section, we illustrate  achievable rates as a function of $n$ per Theorem~\ref{thmMainResultLowerBound}. We do so for $\E[E_1]=P=3\text{dB}$ and various values of $\varepsilon$ and $\Var[E_1]=\E[E_1^2]-P^2$. More specifically, we define
\begin{equation}
R_{n,\varepsilon}^{\text{(EH)}} \triangleq \frac{\frac{n}{2}\log (1+P)- \sqrt{\frac{(2+\varepsilon)n P}{\varepsilon(P+1)}} - n^{\frac{1}{4}}-1} { n + \left\lceil \frac{6\sqrt{a n\log n}}{P} \right\rceil } \label{rValueNumerical}
\end{equation}
(nats per channel use) to be the non-asymptotic rate achievable by save-and-transmit according to \eqref{thmMainResultStatement**}. We plot~$R_{n,\varepsilon}$ against~$n$ in Figure~\ref{figure1} for $\E[E_1]=P=3\text{dB}$ and various values of $\varepsilon$ and $\Var[E_1]$, corresponding to the lines indicated as ``(EH)" respectively.

In order to demonstrate how much the EH constraints \eqref{eqn:eh} degrade the non-asymptotic achievable rates compared to the peak power constraint
\begin{equation}
\Pr\left\{\sum_{k=1}^n X_k^2 \le nP\right\}=1, \label{peakPowerConstraint}
 \end{equation}
 in Figure~\ref{figure1} we also plot the optimal transmission rate~$R_{n, \varepsilon}^{\text{(No-EH)}}$ under the peak power constraint~\eqref{peakPowerConstraint}, corresponding to the lines indicated by ``(No-EH)". Due to Polyanskiy-Poor-Verd\'u \cite[Th.~54, Eq.\ (294)]{PPV10} and Tan-Tomamichel \cite[Th.~1]{TanTom13a},
\begin{align}
R_{n, \varepsilon}^{\text{(No-EH)}} \!=\!\mathrm{C}(P) +\! \sqrt{\frac{\mathrm{V}(P)}{n}} \Phi^{-1} (\varepsilon) +\! \frac{\log n}{2n} + O\!\left(\frac{1}{n}\right)\label{eqn:asymp_expans}
\end{align}
 (nats per channel use) where
\begin{equation}
\mathrm{V}(P) \triangleq  \frac{P(P+2) (\log e)^2}{2(P+1)^2}\quad (\mbox{nats}^2 \mbox{ per channel use})\label{defVP}
\end{equation}
is known as the Gaussian {\em dispersion} function and $\Phi^{-1}$ is the inverse of the cumulative distribution function for the standard Gaussian distribution. We ignore the final correction term in~\eqref{eqn:asymp_expans} when we plot $R_{n, \varepsilon}^{\text{(No-EH)}}$ because it is negligible compared with the first three terms.

In Figure~\ref{figure1}(a), we see that as $\varepsilon$ increases, the backoff of both $R_{n, \varepsilon}^{\text{(EH)}}$ and $R_{n, \varepsilon}^{\text{(No-EH)}}$ from the capacity decreases, which is due to the increase of the magnitude of the second term in \eqref{rValueNumerical}. 
In Figure~\ref{figure1}(b), we see that as $\Var[E_1]$ increases for a fixed $\E[E_1]$, the backoff from the capacity increases, which is due to the explanation in Remark~\ref{remark3} that a longer saving period is required as $\E[E_1^2]$ increases.
As we can see from Figures~\ref{figure1}(a) and~\ref{figure1}(b), the performance degradation due to the EH constraints and the save-and-transmit strategy compared to the peak power constraint is significant.

\section{Save-and-Transmit Strategy} \label{sectionSaveAndTransmit}
In this section, we investigate the save-and-transmit scheme proposed in \cite[Sec.~IV]{ozel12} in the finite blocklength regime. We use this achievability scheme to prove Theorem \ref{thmMainResultLowerBound}.
 \subsection{Prerequisites}
 The following lemma is useful for obtaining a lower bound on the length of the energy-saving phase. The proof is deferred to Appendix~\ref{appendixA}.

\begin{Lemma} \label{lemmaCharacteristicFunction}
Let $m$ and $n$ be two natural numbers. Suppose $\{X_k\}_{k=1}^n$ and $\{E_k\}_{k=1}^{m+n}$ are two sequences of independent and identically distributed (i.i.d.) random variables such that $X^n$ and $E^{m+n}$ are independent\footnote{We note that the random variable $X^n$ here denotes the intended rather than the actual symbols to be sent, and the intended symbols~$X^n$ are chosen to be independent of the EH process $E^{m+n}$. More specifically, in the proof of Theorem~\ref{thmMainResultLowerBound}, we construct the code such that the last $n$ symbols of the intended codeword to be sent over $m+n$ channel uses (which plays the role of $X^n$ in Lemma \ref{lemmaCharacteristicFunction}) are independent of the EH process $E^{m+n}$. However, the actual transmitted codeword is a function of $(X^n, E^{m+n})$ and is correlated with $E^{m+n}$ so that the EH constraints~\eqref{eqn:eh} are satisfied.},
\begin{equation}
\Pr_{p_{E_1}}\{E_1 < 0\}=0, \label{lemmaCharFuncAssump0}
\end{equation}
and
\begin{equation}
\E_{p_{E_1}}[E_1]=\E_{p_{X_1}}[X_1^2]=P. \label{lemmaCharFuncAssump1}
\end{equation}
In addition, suppose there exist two positive numbers $\lambda$ and $a$ such that
\begin{equation}
\max\left\{\E_{p_{E_1}}[E_1^2],\E_{p_{X_1}}\left[X_1^4 e^{\lambda X_1^2}\right]\right\}\le a \, . \label{lemmaCharFuncAssump2}
\end{equation}
If $n\ge 3$ is sufficiently large such that
\begin{equation}
\frac{n}{\log n}\ge \max\left\{ \frac{a}{P^2}\, ,\frac{1}{a\lambda^2} \right\}, \label{lemmaCharFuncAssump3a}
\end{equation}
then we have\footnote{The constant $e^{0.4}$ can be further tightened to $e^{\log n / n}$ by inspecting the proof. Since $e^{0.4}$ and $e^{\log n / n}$ differs by a factor of at most $e^{0.4}$, we prefer stating the upper bound in terms of $e^{0.4}$ to simplify analysis.}
\begin{align}
\Pr_{p_{X^n}p_{E^{m+n}}}\left\{\bigcup_{k=1}^n \left\{\sum_{\ell=1}^k X_\ell^2 \ge \sum_{\ell=1}^{m+k} E_\ell\right\}\right\}  \le  \left(\frac{e^{0.4} }{\log n}\right) e^{2\log n -\frac{mP}{2}\sqrt{\frac{\log n}{a n}}}.
\end{align}
\end{Lemma}
\begin{Corollary}\label{corollaryCharFunc}
 Suppose $\{X_k\}_{k=1}^n$ and $\{E_k\}_{k=1}^{m+n}$ are two sequences of independent and identically distributed (i.i.d.) random variables such that $X^n$ and $E^{m+n}$ are independent and also satisfy \eqref{lemmaCharFuncAssump0} and \eqref{lemmaCharFuncAssump1}. If $X_1$ is a Gaussian random variable, then by setting $a\triangleq \max\left\{\E_{p_{E_1}}[E_1^2], 12\sqrt{2} P^2\right\}$ we have
\begin{align}
\Pr_{p_{X^n}p_{E^{m+n}}}\left\{\bigcup_{k=1}^n \left\{\sum_{\ell=1}^k X_\ell^2 \ge \sum_{\ell=1}^{m+k} E_\ell\right\}\right\} \le  \left(\frac{e^{0.4} }{\log n}\right) e^{2\log n -\frac{mP}{2}\sqrt{\frac{\log n}{a n}}} \label{corollaryEqn}
\end{align}
for all sufficiently large~$n$ that satisfies
\begin{equation*}
\frac{n}{\log n} \ge \max \left\{\frac{\E_{p_{E_1}}[E_1^2]}{P^2},12\sqrt{2}\right\}.
\end{equation*}
\end{Corollary}
\begin{IEEEproof}
Suppose $X_1$ is a zero-mean Gaussian random variable with variance~$P$. Then, straightforward calculation using \eqref{eqnNormalDist} reveals that
\begin{equation}
\E_{p_{X_1}}\left[X_1^4 e^{\frac{X_1^2}{4P}}\right] = 12\sqrt{2} P^2. \label{corollaryEq1InProof}
\end{equation}
Fix
\begin{equation}
\lambda\triangleq\frac{1}{4P} \label{fixLambda}
\end{equation}
 and fix an
\begin{align}
a& \triangleq \max\left\{\E_{p_{E_1}}[E_1^2],\E_{p_{X_1}}\left[X_1^4 e^{\lambda X_1^2}\right]\right\} \notag\\
& \stackrel{\text{(a)}}{=} \max\left\{\E_{p_{E_1}}[E_1^2],12\sqrt{2} P^2\right\} \label{defAinCorollaryProof}
\end{align}
where (a) follows \eqref{corollaryEq1InProof} and \eqref{fixLambda}. It then follows from Lemma~\ref{lemmaCharacteristicFunction} that
\eqref{corollaryEqn} holds
for all sufficiently large~$n\ge 3$ such that
\begin{align}
\frac{n}{\log n}
& \ge \max\left\{ \frac{a}{P^2}\, , \frac{1}{a\lambda^2}\right\} \notag\\
& \stackrel{\eqref{fixLambda}}{=} \max\left\{ \frac{a}{P^2}\, , \frac{16P^2}{a}\right\} \notag\\
&\stackrel{\eqref{defAinCorollaryProof}}{=}  \max \left\{\frac{\E_{p_{E_1}}[E_1^2]}{P^2}, 12\sqrt{2} , \min\left\{\frac{16 P^2}{12\sqrt{2}P^2}, \frac{16P^2}{\E_{p_{E_1}}[E_1^2]}\right\}\right\} \notag\\
& \stackrel{\text{(a)}}{=} \max \left\{\frac{\E_{p_{E_1}}[E_1^2]}{P^2}, 12\sqrt{2}\right\},
\end{align}
where (a) follows from the fact that $\min\left\{\frac{16 P^2}{12\sqrt{2}P^2}, \frac{16P^2}{\E_{p_{E_1}}[E_1^2]}\right\} \le 12\sqrt{2}$.
\end{IEEEproof}
The following lemma~\cite{sha57} is standard for proving achievability results in the finite blocklength regime and its proof can be found in \cite[Th.~3.8.1]{Han10}.
\begin{Lemma}[Implied by Shannon's bound~\cite{sha57}] \label{lemmaFeinstein}
Let $p_{X^n, Y^n}$ be the probability distribution of a pair of random variables $(X^n,Y^n)$. Let $\{X^n(i), Y^n(i)\}_{i=1}^{\infty}$ be a sequence of independent  random variables where $(X^n(1), Y^n(1))$ is distributed according to $p_{X^n, Y^n}$.
 For each $\delta>0$ and each~$M\in \mathbb{N}$, we have
\begin{align}
\Pr\bigg\{ \bigcup_{i=2}^M\left\{\log\left(\frac{p_{Y^n|X^n}(Y^n(1)|X^n(i))}{p_{Y^n}(Y^n(1))} \right)> \log M + n\delta \right\}\bigg\} \le e^{-n\delta}.
\end{align}
\end{Lemma}
\subsection{Proof of Theorem~\ref{thmMainResultLowerBound}} \label{sectionProofOfMainResult}
Fix an $\varepsilon\in(0,1)$. Define
 \begin{equation}
a\triangleq \max\left\{\E_{p_{E_1}}[E_1^2], 12\sqrt{2} P^2\right\}. \label{defAinProof}
\end{equation}
 Fix a sufficiently large~$n\ge 3$ such that \eqref{sufficientlyLargeNinProof}, \eqref{sufficientlyLargeNinProof*} and \eqref{sufficientlyLargeNinProof**} hold. Define
\begin{equation}
m\triangleq \left\lceil \frac{6\sqrt{a n\log n}}{P} \right\rceil, \label{defm}
\end{equation}
which specifies the number of time slots which are used for saving energy.
Consider the random code that uses the channel $m+n$ times as follows: \\
\textbf{Save-and-Transmit Random Codebook Construction}\\
Let $\mathbf{0}^m$ denote the length-$m$ zero tuple. Let $p_X$ be the distribution of a zero-mean Gaussian random variable $X$ whose variance is~$P$, i.e.,
 \begin{equation}
 p_{X}(x) \triangleq \mathcal{N}(x; 0, P) \label{defDistX}.
 \end{equation}
 In addition, let $p_{X^n}$ be the product distribution of the~$n$ independent copies of~$X$. Construct~$M$ i.i.d.\ random tuples denoted by $X^n(1), X^n(2), \ldots, X^n(M)$ such that $X^n(1)$ is distributed according to~$p_{X^n}$, where $M$ will be carefully chosen later when we evaluate the probability of decoding error.
 Define
 \begin{equation}
 \tilde X^{m+n}(i)\triangleq (\mathbf{0}^m, X^n(i)) \label{defTildeXmn}
 \end{equation}
  for each $i\in\{1, 2, \ldots, M\}$ and construct the random codebook
\begin{equation}
\big\{\tilde X^{m+n}(i) \,\big|\, i\in\{1, 2, \ldots, M\} \big\}. \label{defCodebook}
\end{equation}
The codebook is revealed to both the encoder and the decoder.
To facilitate discussion, we let $X_k(i)$ and $\tilde X_k(i)$ denote the $k^{\text{th}}$ symbols in $X^n(i)$ and $\tilde X^{m+n}(i)$ respectively for each $i$.
Since the first~$m$ symbols of each random codeword $\tilde X^{m+n}(i)$ are zeros by \eqref{defTildeXmn}, the source will just transmit~$0$ with probability one until time slot $m+1$ when the amount of energy $\sum_{k=1}^{m+1} E_k$ is available for encoding $\tilde X_{m+1}(W)\stackrel{\eqref{defTildeXmn}}{=}X_1(W)$.
\\
\textbf{Encoding under  the EH Constraints}\\
The source has the knowledge of $E^k$ before transmitting its symbol in time slot~$k$ for each $k\in\{1, 2, \ldots, m+n\}$. For each $i\in\{1, 2, \ldots, M\}$, recalling that $\tilde X_k(i)$ is the $k^{\text{th}}$ element of $\tilde X^{m+n}(i)\stackrel{\eqref{defTildeXmn}}{=} (\mathbf{0}^m, X^n(i))$, we construct recursively for $k=1, 2, \ldots, m+n$ the random variable
\begin{align}
\hat X_k(i, E^k)\triangleq
\begin{cases}
\tilde X_k(i) &  \text{if $(\tilde X_k(i))^2 \le \sum\limits_{\ell=1}^k E_\ell - \sum\limits_{\ell=1}^{k-1}(\hat X_\ell(i, E^{\ell}))^2$,}\\
 0 & \text{otherwise.}
\end{cases} \label{defHatXk}
\end{align}
To send message~$W$ which is uniformly distributed on $\{1, 2, \ldots, M\}$, the source transmits $\hat X_k(W, E^k)$ in time slot~$k$ for each $k\in\{1, 2, \ldots, m+n\}$. Note that the source transmits~$0$ with probability one in the first~$m$ times slots by \eqref{defTildeXmn} and \eqref{defHatXk}, and the transmitted codeword $(\hat X_1(W, E^1), \hat X_2(W, E^2), \ldots, \hat X_{m+n}(W, E^{m+n}))$ satisfies the EH constraints \eqref{eqn:eh} by \eqref{defHatXk}.
\\
\textbf{Threshold Decoding}\\
Upon receiving
\begin{equation}
\hat Y^{m+n} =\hat X^{m+n}(W, E^{m+n}) + Z^{m+n} \label{defHatYmn}
\end{equation}
where
\begin{align}
\hat X^{m+n}(W, E^{m+n}) \triangleq(\hat X_1(W, E^1), \hat X_2(W, E^2), \ldots, \hat X_{m+n}(W, E^{m+n})) \label{defHatXmn}
 \end{align}
 denotes the transmitted tuple specified in \eqref{defHatXk} and $Z^{m+n}$ is a tuple consisting of i.i.d.\ standard normal random variables by the channel law (cf.\ \eqref{defChannelInDefinition*}), the destination constructs its subtuple denoted by $\bar Y^n$ by keeping only the last~$n$ symbols of $\hat Y^{m+n}$. Recalling that $q_{Y|X}$ denotes the channel law and $p_X$ was chosen in the above codebook construction to be the distribution of the zero-mean Gaussian random variable with variance~$P$, we define the joint distribution
 \begin{equation}
p_{X,Y}\triangleq p_Xq_{Y|X}, \label{defJointDistribution}
 \end{equation}
and define $p_{X^n, Y^n}$ to be the probability distribution of~$n$ independent copies of $(X,Y)$ distributed according to $p_{X,Y}$, i.e.,
 \begin{equation}
 p_{X^n, Y^n}(x^n, y^n) \triangleq \prod_{k=1}^n p_{X,Y}(x_k,y_k) \label{factorizationOfPXY}
 \end{equation}
 for all $(x^n, y^n)\in\mathbb{R}^2$. Then, the decoder declares $\varphi(\bar Y^n) \in\{1, 2, \ldots, M\}$ (with a slight abuse of notation, we write $\varphi(\bar Y^n)$ instead of $\varphi(\hat Y^{m+n})$) to be the transmitted message where $\varphi(\bar Y^{n})$ is the decoding function defined as follows:
 If there exists a unique index~$j$ such that
\begin{equation}
\log\left(\frac{p_{Y^n|X^n}(\bar Y^n| X^n(j))}{p_{Y^n}(\bar Y^n)}\right) > \log M + n^{\frac{1}{4}}, \label{decodingRule}
\end{equation}
then $\varphi(\bar Y^{n})$ is assigned the value~$j$.
Otherwise, $\varphi(\bar Y^{n})$ is assigned a random value uniformly distributed on $\{1,2, \ldots,M\}$.
\\
\textbf{Calculating the Probability of Violating the EH Constraints}\\
It follows from \eqref{defHatXmn} and \eqref{defHatXk} that
\begin{align}
\Pr\left\{\hat X^{m+n}(W, E^{m+n}) = \tilde X^{m+n}(W) \left| \bigcap_{k=1}^{m+n}\left\{ \sum_{\ell=1}^{k}(\tilde X_{\ell}(W))^2 \le \sum_{\ell=1}^{k} E_\ell \right\}\right.\right\}=1. \label{eqn1InCalculationErrorProb} 
\end{align}
Defining $\bar X^n(W, E^{m+n})$ to be the tuple containing the last~$n$ symbols of $\hat X^{m+n}(W, E^{m+n})$, we obtain from \eqref{eqn1InCalculationErrorProb} and \eqref{defTildeXmn} that
\begin{align}
&\Pr\left\{\bar X^n(W, E^{m+n}) = X^{n}(W)\left| \bigcap_{k=1}^{n}\left\{ \sum_{\ell=1}^{k}(X_{\ell}(W))^2 \le \sum_{\ell=1}^{m+k} E_\ell \right\}\right.\right\}=1. \label{eqn2InCalculationErrorProb*} 
\end{align}
%
%
Combining Corollary~\ref{corollaryCharFunc}, \eqref{defAinProof} and \eqref{sufficientlyLargeNinProof} and noting that $E^{m+n}$ and $(W,X^n(W))$ are independent by construction, we obtain
\begin{align}
\Pr\left\{\bigcup_{k=1}^n \left\{\sum_{\ell=1}^k (X_\ell(W))^2 > \sum_{\ell=1}^{m+k} E_\ell\right\}\right\} \le  \left(\frac{e^{0.4} }{\log n}\right) e^{2\log n -\frac{mP}{2}\sqrt{\frac{\log n}{a n}}}  \label{eqn4InCalculationErrorProb},
\end{align}
which implies that
\begin{align}
\Pr\left\{\bigcap_{k=1}^n \left\{\sum_{\ell=1}^k (X_\ell(W))^2 \le \sum_{\ell=1}^{m+k} E_\ell\right\}\right\}
&\ge  1- \left(\frac{e^{0.4} }{\log n}\right) e^{2\log n -\frac{mP}{2}\sqrt{\frac{\log n}{a n}}}\notag\\*
& \stackrel{\eqref{defm}}{\ge} 1- \frac{e^{0.4} }{n\log n} \notag\\
& \stackrel{\eqref{sufficientlyLargeNinProof**}}{\ge} 1-\frac{\varepsilon}{2+\varepsilon}\,. \label{eqn5InCalculationErrorProb}
\end{align}
Using \eqref{eqn2InCalculationErrorProb*} and \eqref{eqn5InCalculationErrorProb}, we have
\begin{equation}
\Pr\left\{ \bar X^n(W, E^{m+n}) = X^{n}(W)\right\} \ge 1- \frac{\varepsilon}{2+\varepsilon}\,. \label{eqn6InCalculationErrorProb*}
\end{equation}
\textbf{Calculating the Probability of Decoding Error}\\
Defining $\bar Z^n$ to be the tuple containing the last~$n$ symbols of $Z^{m+n}$ and recalling $\bar X^n(W, E^{m+n})$ and $\bar Y^n$ are the tuples containing the last~$n$ symbols of $\hat X^{m+n}(W, E^{m+n})$ and $\hat Y^{m+n}$ respectively, we obtain from \eqref{defHatYmn} and \eqref{eqn6InCalculationErrorProb*} that
\begin{equation}
\Pr\left\{ \bar Y^n = X^{n}(W) + \bar Z^n \right\} \ge 1-  \frac{\varepsilon}{2+\varepsilon}\,, \label{eqn6InCalculationErrorProb}
\end{equation}
where $X^n(W)$ and $\bar Z^n$ are independent and $\bar Z^n$ consists of i.i.d.\ standard normal random variables by the channel law.
Following \eqref{decodingRule} and \eqref{eqn6InCalculationErrorProb}, we define the events\footnote{The term $n^{\frac{1}{4}}$ corresponds to the choice of $n\delta = n^{\frac{1}{4}}$ in Lemma~\ref{lemmaFeinstein}, which can be replaced by any function of the order $o(\sqrt{n \log n})$ without affecting the achievable second-order term in~\eqref{thmMainResultStatement} (cf.\ \eqref{defChoiseOfM}). We do not optimize the choice of $n\delta$ because it affects only higher-order terms which are negligible compared with the first- and second-order terms.}
 \begin{equation}
 \mathcal{E}_{i|w} \!\triangleq\! \left\{ \log\!\left(\frac{p_{Y^n|X^n}(X^n(w) \!+\! \bar Z^n| X^n(i))}{p_{Y^n}(X^n(w) + \bar Z^n)}\right) \!\le\! \log M \!+\! n^{\frac{1}{4}} \right\} \label{defEiw}
 \end{equation}
and consider the following chain of inequalities for each $w\in \{1, 2, \ldots, M\}$:
\begin{align}
& \Pr_{p_W(\prod_{i=1}^M p_{X^n(i)})p_{\bar Z^n}}\!\!\left\{\!\left.\mathcal{E}_{w|w} \cup \bigcup_{j\in \{1,2, \ldots, M\}\setminus\{w\}}  \mathcal{E}_{j|w}^c\right|\!W\!=\!w\right\} \notag \\
& \stackrel{\text{(a)}}{=} \Pr_{p_W(\prod_{i=1}^M p_{X^n(i)})p_{\bar Z^n}}\left\{\left.\parbox[c]{1 in}{$\mathcal{E}_{1|1} \cup \bigcup_{j=2}^M  \mathcal{E}_{j|1}^c$}\,\right|W=1\right\}\notag\\
&\le \Pr_{p_W(\prod_{i=1}^M p_{X^n(i)})p_{\bar Z^n}}\left\{\left.\mathcal{E}_{1|1} \right|W=1\right\} + \Pr_{p_W(\prod_{i=1}^M p_{X^n(i)})p_{\bar Z^n}}\left\{\left.\cup_{j=2}^M \, \mathcal{E}_{j|1}^c\,\right|W=1\right\}\notag\\
& \stackrel{\text{(b)}}{\le} \Pr_{p_Wp_{X^n(1)}p_{\bar Z^n}}\left\{\left.  \mathcal{E}_{1|1} \right|W=1\right\} + e^{-n^{1/4}}\notag\\
& =\Pr_{p_{X^n(1)}p_{\bar Z^n}}\left\{\mathcal{E}_{1|1}\right\} + e^{-n^{1/4}}\notag\\
& \stackrel{\text{(c)}}{=}\Pr_{\prod_{k=1}^n p_{X_k(1)}p_{\bar Z_k}}\left\{\mathcal{E}_{1|1}\right\} + e^{-n^{1/4}}\notag\\
&\stackrel{\eqref{sufficientlyLargeNinProof*}}{\le}\Pr_{\prod_{k=1}^n p_{X_k(1)}p_{\bar Z_k}}\left\{\mathcal{E}_{1|1}\right\}+ \frac{\varepsilon^2}{2+\varepsilon}\, , \label{eqn6*InCalculationErrorProb}
\end{align}
where
\begin{enumerate}
\item[(a)] follows from symmetry of the random codebook construction.
\item[(b)] follows from Lemma~\ref{lemmaFeinstein} and \eqref{defEiw}.
\item[(c)] follows from the fact that that $X^n(1)$ and $\tilde Z^n$ are i.i.d.\ copies of $X_1(1)$ and $\tilde Z_1$ respectively by construction.
\end{enumerate}
In order to ensure the first term in \eqref{eqn6*InCalculationErrorProb} can be upper bounded by a simple term, we choose $M$ to be the unique integer that satisfies
\begin{align}
\log (M+1)& \ge n \E_{p_{X,Y}}\left[\log\left(\frac{p_{Y|X}(Y|X)}{p_{Y}(Y)}\right)\right]- \sqrt{\frac{(2+\varepsilon)n}{\varepsilon} \Var_{p_{X,Y}}\left[\log\left(\frac{p_{Y|X}(Y|X)}{p_{Y}(Y)}\right)\right]} - n^{\frac{1}{4}} \notag\\
&> \log M. \label{defChoiseOfM}
\end{align}
Following \eqref{eqn6*InCalculationErrorProb}, we consider the following chain of inequalities where the random variables are distributed according to $\prod_{k=1}^n p_{X_k(1)}p_{\bar Z_k}$:
\begin{align}
\Pr\left\{\mathcal{E}_{1|1}\right\}
&\stackrel{\eqref{defEiw}}{=}\Pr\left\{\sum_{k=1}^n \log\left(\frac{p_{Y|X}(X_k(1) + \bar Z_k| X_k(1))}{p_{Y}(X_k(1) + \bar Z_k)}\right) \le \log M + n^{\frac{1}{4}}\right\}  \notag\\
&\stackrel{\eqref{defChoiseOfM}}{\le} \Pr\left\{\parbox[c]{4.5 in}{$\sum_{k=1}^n \log\left(\frac{p_{Y|X}(X_k(1) + \bar Z_k| X_k(1))}{p_{Y}(X_k(1) + \bar Z_k)}\right)\\\le n\E_{p_{X,Y}}\left[\log\left(\frac{p_{Y|X}(Y|X)}{p_{Y}(Y)}\right)\right] - \sqrt{\frac{(2+\varepsilon)n}{\varepsilon} \Var_{p_{X,Y}}\left[\log\left(\frac{p_{Y|X}(Y|X)}{p_{Y}(Y)}\right)\right]}$}\right\}  \notag\\
&\stackrel{\text{(a)}}{\le} \frac{\varepsilon}{2+\varepsilon}\, , \label{eqn6**InCalculationErrorProb}
\end{align}
where (a) follows from Chebyshev's inequality and the facts based on \eqref{defJointDistribution}, \eqref{defDistX} and \eqref{defChannelInDefinition*} that $(X_k(1), X_k(1)+\bar Z_k)$ and $(X,Y)$ have the same distribution for each $k\in\{1, 2, \ldots, n\}$.
Combining \eqref{eqn6*InCalculationErrorProb} and \eqref{eqn6**InCalculationErrorProb}, we obtain
\begin{align}
 \Pr_{p_W(\prod_{i=1}^M p_{X^n(i)})p_{\bar Z^n}}\!\!\left\{\!\left.\mathcal{E}_{w|w} \cup \bigcup_{j\in \{1,2, \ldots, M\}\setminus\{w\}}  \mathcal{E}_{j|w}^c\right|\!W\!=\!w\right\}
\le \frac{\varepsilon + \varepsilon^2}{2+\varepsilon} \label{eqn7InCalculationErrorProb}
\end{align}
for each $w\in \{1, 2, \ldots, M\}$.
We are ready to compute the probability of decoding error as follows, where the random variables are distributed according to~$p_{W, X^n(W)}p_{\bar Z^n}p_{\bar Y^n|W, X^n(W), \bar Z^n}$:
\begin{align}
 \Pr\left\{\varphi(\bar Y^{n})\ne W \right\}
& \stackrel{\eqref{eqn6InCalculationErrorProb}}{\le}  \Pr\left\{\left\{\varphi(\bar Y^{n})\ne W\right\}\cap \{\bar Y^n=X^n(W) + \bar Z^n\} \right\} + \frac{\varepsilon}{2+\varepsilon} \notag\\
& \le \Pr\left\{\varphi(X^n(W) + \bar Z^n)\ne W \right\} + \frac{\varepsilon}{2+\varepsilon} \notag\\
& \stackrel{\text{(a)}}{\le}\varepsilon \label{eqn8InCalculationErrorProb}
\end{align}
where (a) follows from the threshold decoding rule (cf.\ \eqref{decodingRule} and \eqref{defEiw}) and \eqref{eqn7InCalculationErrorProb}. Using \eqref{defm}, \eqref{defCodebook}, \eqref{defChoiseOfM} and \eqref{eqn8InCalculationErrorProb}, we conclude that the constructed code is an $(n+ m, M, \varepsilon)$-code that satisfies
\begin{align}
\log (M+1)\ge n \E_{p_{X,Y}}\left[\log\left(\frac{p_{Y|X}(Y|X)}{p_{Y}(Y)}\right)\right] - \sqrt{\frac{(2+\varepsilon)n}{\varepsilon} \Var_{p_{X,Y}}\left[\log\left(\frac{p_{Y|X}(Y|X)}{p_{Y}(Y)}\right)\right]} - n^{\frac{1}{4}},\label{eqn9InCalculationErrorProb}
\end{align}
which implies from \eqref{defJointDistribution}, \eqref{defDistX} and \eqref{defChannelInDefinition*} that
\begin{equation}
\log (M+1) \ge \frac{n}{2}\log (1+P)- \sqrt{\frac{(2+\varepsilon)nP}{\varepsilon(P+1)}} - n^{\frac{1}{4}}, \label{eqn10InCalculationErrorProb}
\end{equation}
which then implies that
\begin{align}
\log M &\ge \frac{n}{2}\log (1+P)- \sqrt{\frac{(2+\varepsilon)nP}{\varepsilon(P+1)}} - n^{\frac{1}{4}}-1 \label{eqn11InCalculationErrorProb**}\\
& \ge \frac{(n+m)-m}{2}\log (1+P)- \sqrt{\frac{(2+\varepsilon)(m+n)P}{\varepsilon(P+1)}} - (m+n)^{\frac{1}{4}}-1. \label{eqn11InCalculationErrorProb}
\end{align}
Equation~\eqref{thmMainResultStatement**} then follows from \eqref{eqn11InCalculationErrorProb**}, \eqref{defAinProof}
and \eqref{defm}. It remains to prove~\eqref{thmMainResultStatement}.
Since
$
m \ge 0 $
and
$
 m \le \frac{6\sqrt{an\log n}}{P} + 1$
by \eqref{defm}, it follows that
\begin{align}
\frac{6\sqrt{a(m+n)\log(m+n)}}{P}+1 \!\ge\! \frac{6\sqrt{a n\log n}}{P}\!\!+\!1 \ge\! m,
\end{align}
which implies from \eqref{eqn11InCalculationErrorProb} that
\begin{align}
\log M
& \ge \frac{m\!+\!n}{2}\log (1\!+\!P)- \frac{3\sqrt{a(m\!+\!n)\log(m\!+\!n)}}{P}\log (1\!+\!P) \notag\\*
&\quad - \sqrt{\frac{(2+\varepsilon)(m+n)P}{\varepsilon(P+1)}} - (m+n)^{\frac{1}{4}}-\frac{1}{2}\log(1+P)-1 \, . \label{eqn12InCalculationErrorProb}
\end{align}
Equation~\eqref{thmMainResultStatement} then follows from \eqref{eqn12InCalculationErrorProb}, \eqref{defAinProof} 
and \eqref{defm} by letting $n^*\triangleq m+n$.

\section{Discrete Memoryless Energy-Harvesting Channel} \label{sectionFiniteAlphabet}
\subsection{Channel Model and Main Result}
We now consider a discrete memoryless EH (DM-EH) channel which consists of a finite input alphabet denoted by~$\mathcal{X}$, a finite output alphabet denoted by~$\mathcal{Y}$, a transition matrix $q_{Y|X}$ and a cost function $c:\mathcal{X}\rightarrow \mathbb{R}_+$. We assume that $\mathcal{X}$ contains a symbol denoted by $0$ such that $c(0)=0$, where $0$ represents the idle symbol that consumes no energy.

As an example of a cost function, let us consider a binary alphabet $\mathcal{X}=\{0,1\}$ and $c(x) = x$ for all $x\in\{0,1\}$. The cost function~$c$ characterizes the cost of sending a symbol $x\in \mathcal{X}$. For a length-$n$ sequence~$x^n$, the total cost is $\sum_{k=1}^n c(x_k)$, which is equivalent to the \emph{weight} of~$x^n$ (number of ones in~$x^n$). Since typically we need to expend some energy to transmit~1 while no energy is required to transmit~0 (the transmitter stays ``silent"), $\sum_{k=1}^n c(x_k)$ is a reasonable measure of energy consumption for transmitting~$x^n$.

 At each discrete time $k \in \{1,2,\ldots\}$, suppose a random amount of energy $E_k\in [0, \infty)$ arrives at the buffer and the source~$\mathrm{s}$ transmits $X_k\in \mathcal{X}$ such that
 \begin{align}
 \Pr\left\{c(X_k) \le E_k + \sum_{\ell=1}^{k-1}(E_\ell -c(X_\ell))\right\}= \Pr\left\{\sum_{\ell=1}^k c(X_\ell) \le \sum_{\ell=1}^k E_\ell \right\}=1.  \label{EHResultIntroDMC}
 \end{align}
 We assume that $E_1, E_2, \ldots$ are independent and identically distributed (i.i.d.) non-negative random variables, where $\E[E_1]=P$ and $\E[E_1^2]<+\infty$. The destination~$\mathrm{d}$ receives $Y_k$ from the channel output in time slot~$k$ for each $k\in\{1, 2, \ldots\}$, where $p_{Y_k|X_k}$ is distributed according to the channel law such that $p_{Y_k|X_k}(y_k|x_k)=q_{Y|X}(y_k|x_k)$ for all $(x_k, y_k)\in \mathcal{X}\times \mathcal{Y}$.
   After~$n$ time slots, node~$\mathrm{d}$ declares~$\hat W$ to be the transmitted~$W$ based on $Y^n$. We formally define a code for the DM-EH channel as follows.
\begin{Definition} \label{defCodeDMC}
An {\em $(n, M)$-code} consists of the following:
\begin{enumerate}
\item A message set
$
\mathcal{W}\triangleq \{1, 2, \ldots, M\}
$
 at node~$\mathrm{s}$. Message $W$ is uniform on $\mathcal{W}$.

\item  A sequence of  encoding functions
$
f_k : \mathcal{W}\times \mathbb{R}_+^{k}\rightarrow \mathcal{X}
$
such that
$
X_k=f_k (W, E^{k})
$
and \eqref{EHResultIntroDMC} holds.
\item A decoding function
$
\varphi :
\mathcal{Y}^{n} \rightarrow \mathcal{W},
$
where $\varphi$ is the decoding function for $W$ at node~$\mathrm{d}$ such that
$
 \hat W = \varphi(Y^{n})$.
\end{enumerate}
\end{Definition}
\begin{Definition}\label{defDMchannel}
The \textit{discrete memoryless EH} (DM-EH) channel is characterized by $q_{Y|X}$ such that the following holds for any $(n, M)$-code: For each $k\in\{1, 2, \ldots, n\}$,
\begin{align}
p_{W, E^k, X^k, Y^k}
 = p_{W, E^k, X^k, Y^{k-1}}p_{Y_k|X_k} \label{memorylessStatement*DM}
\end{align}
where
\begin{equation}
p_{Y_k|X_k}(y_k|x_k) = q_{Y|X}(y_k|x_k) \label{defChannelInDefinition*DM}
\end{equation}
for all $(x_k, y_k)\in \mathcal{X}\times \mathcal{Y}$.
Since $p_{Y_k|X_k}$ does not depend on~$k$ by \eqref{defChannelInDefinition*DM}, the channel is stationary.
\end{Definition}

 For any $(n, M)$-code defined on the DM-EH channel, let $p_{W,E^n, X^n, Y^n, \hat W}$ be the joint distribution induced by the code. Similar to \eqref{memorylessStatement}, we can factorize $p_{W,E^n, X^n, Y^n, \hat W}$ as
\begin{align}
 p_{W,E^n, X^n, Y^n, \hat W}
 = p_W \left(\prod_{k=1}^n p_{E_k} p_{X_k|W, E^k} p_{Y_k|X_k}\right)p_{\hat W |Y^n}, \label{memorylessStatementDM}
\end{align}
which implies through straightforward calculations that
\begin{align}
 p_{W, E^n, X^n, Y^n}
 = p_{W, E^n, X^n}\prod_{k=1}^n p_{Y_k|X_k}. \label{memorylessStatementDM^*}
\end{align}

\begin{Definition} \label{defErrorProbabilityDMC}
For an $(n, M)$-code defined on the DM-EH channel, we can calculate according to \eqref{memorylessStatementDM} the \textit{average probability of decoding error} defined as $\Pr\big\{\hat W \ne W\big\}$.
We call an $(n, M)$-code with average probability of decoding error no larger than $\varepsilon$ an $(n, M, \varepsilon)$-code.
\end{Definition}
\medskip
For each $\varepsilon\in[0,1)$, we define the $\varepsilon$-achievable rate and the $\varepsilon$-capacity in the same way as done in Definition~\ref{defAchievableRate} and Definition~\ref{defCapacity}.
%
Define the capacity-cost function
\begin{align}
\mathrm{C}(P)\triangleq \max\limits_{p_X: \E_{p_X}[c(X)]=P}\!\E_{p_Xq_{Y|X}} \!\! \left[\log \! \left(\frac{q_{Y|X}(Y|X)}{\sum_{x\in \mathcal{X} }p_{X}(x)q_{Y|X}(Y|x)}\right)\right]= \max\limits_{p_X: \E_{p_X}[c(X)]=P}I_{p_X q_{Y|X}}(X;Y) \label{defCDMC}  .
\end{align}
Note that this is the capacity of the DMC when each codeword $X^n$ has to satisfy the cost constraint $\sum_{\ell=1}^n c(X_\ell)\le nP$.
Similarly, define
\begin{align}
\mathrm{V}(P)\triangleq \max\limits_{p_X: \E_{p_X}[c(X)]=P}\!\! \Var_{p_Xq_{Y|X}}\!\! \left[\log \! \left(\frac{q_{Y|X}(Y|X)}{\sum_{x\in \mathcal{X} }p_{X}(x)q_{Y|X}(Y|x)}\right)\right]. \label{defVDMC}
\end{align}
We also note that this is the maximum variance of the log-likelihood ratio of the channel and the output distribution where the input distribution is constrained to be such that the same cost constraint is satisfied.
 The following theorem is the main result in this section.
\begin{Theorem}\label{thmMainResultLowerBoundDMC}
Let $\varepsilon\in(0,1)$, and define
\begin{equation}
a\triangleq \max\left\{\E_{p_{E_1}}[E_1^2], \max\left\{\left. (c(x))^2 e^{c(x)}\right| x\in \mathcal{X}\right\} \right\}.
\end{equation}
For each sufficiently large integer~$n\ge 3$ such that
 \begin{equation}
 \frac{n}{\log n} \ge \frac{a}{P^2}, \label{sufficientlyLargeNinProofDMC}
 \end{equation}
 \begin{equation}  n\ge \left(\log\left(\frac{2+\varepsilon}{\varepsilon^2}\right)\right)^4 \label{sufficientlyLargeNinProof*DMC}
  \end{equation}
  and
   \begin{equation}
 n\log n \ge \frac{e^{0.4}(2+\varepsilon)}{\varepsilon},\label{sufficientlyLargeNinProof**DMC}
  \end{equation}
there exists an~$(n^*, M, \varepsilon)$-code with $n^* \triangleq n + \left\lceil \frac{6\sqrt{a n\log n}}{P} \right\rceil$ such that
\begin{align}
\log M &\ge n^*\mathrm{C}(P) - \frac{6 \mathrm{C}(P) \sqrt{a n^*\log n^*}}{P} - \sqrt{\frac{(2+\varepsilon)n^* \mathrm{V}(P)}{\varepsilon}} \notag\\
 &\quad- (n^*)^{\frac{1}{4}}-\mathrm{C}(P)-1\,.
\end{align}
\end{Theorem}
\begin{Remark}
Since $\frac{1}{n^*}\sum_{k=1}^{n^*} E_k$ converges to $\E_{p_{E_1}}\left[E_1\right]=P$ with probability one by the strong law of large numbers, it follows from the power constraint \eqref{EHResultIntroDMC} and the strong converse theorem for the DMC with cost constraint \cite{Wolfowitz,kost15} that the $\varepsilon$-capacity of the DM-EH channel is upper bounded by $\mathrm{C}(P)$, defined in \eqref{defCDMC}. Therefore, Theorem~\ref{thmMainResultLowerBoundDMC} shows that the $\varepsilon$-capacity of the DM-EH channel is $\mathrm{C}(P)$. In addition, Theorem~\ref{thmMainResultLowerBoundDMC} implies a lower bound on the second-order term that is proportional to $-\sqrt{{n^*}\log {n^*}}\,$.
\end{Remark}

\begin{Remark}
We observe that $a\ge\E[ E_1^2] \ge \E [ E_1 ]^2=P^2$. Thus as $P$ increases, the magnitude of the coefficient of the second-order term (that scales as $\sqrt{n^*  \log n^*}$ and depends on the length of the saving period alone due to similar reasons provided for the Gaussian case in Remark~\ref{remark3}) increases with $P$, which implies that a longer saving period is needed to maintain a fixed probability of outage (the term ``outage" is as described in Remark~\ref{remark3}). Similarly, as~$\E[E_1^2]$ increases while~$P$ is fixed, the variance of the energy arrival process increases and hence a longer saving period is needed to maintain a certain fixed outage probability. 
\end{Remark}
\subsection{Save-and-Transmit Strategy}
Before we provide the proof of Theorem~\ref{thmMainResultLowerBoundDMC}, we state the following useful lemma that is analogous to Lemma~\ref{lemmaCharacteristicFunction}.
\begin{Lemma} \label{lemmaCharacteristicFunctionDMC}
Let $m$ and $n$ be two natural numbers. Suppose $\{X_k\}_{k=1}^n$ and $\{E_k\}_{k=1}^{m+n}$ are two sequences of independent and identically distributed (i.i.d.) random variables such that $X^n$ and $E^{m+n}$ are independent,
\begin{equation}
\Pr_{p_{E_1}}\{E_1 < 0\}=0, \label{lemmaCharFuncAssump0DMC}
\end{equation}
and
\begin{equation}
\E_{p_{E_1}}[E_1]=\E_{p_{X_1}}[c(X_1)]=P. \label{lemmaCharFuncAssump1DMC}
\end{equation}
In addition, suppose there exists an $a>0$ such that
\begin{equation}
\max\left\{\E_{p_{E_1}}[E_1^2],\max\left\{\left.(c(x))^2 e^{c(x)}\right| x\in \mathcal{X} \right\}\right\}\le a \, . \label{lemmaCharFuncAssump2DMC}
\end{equation}
If $n\ge 3$ is sufficiently large such that
\begin{equation}
\frac{n}{\log n}\ge \frac{a}{P^2}\, , \label{lemmaCharFuncAssump3aDMC}
\end{equation}
then we have
\begin{align}
\Pr_{p_{X^n}p_{E^{m+n}}}\left\{\bigcup_{k=1}^n \left\{\sum_{\ell=1}^k X_\ell^2 \ge \sum_{\ell=1}^{m+k} E_\ell\right\}\right\} \le  \left(\frac{e^{0.4} }{\log n}\right) e^{2\log n -\frac{mP}{2}\sqrt{\frac{\log n}{a n}}}.
\end{align}
\end{Lemma}
\begin{IEEEproof}
If we replace every instance of $X_k^2$ with $c(X_k)$ for each $k\in\{1, 2, \ldots, n\}$ and set $\lambda\triangleq 1$ in the proof of Lemma~\ref{lemmaCharacteristicFunction} in Appendix~\ref{appendixA}, then the resultant proof will immediately lead  to this lemma.
\end{IEEEproof}
We are ready to prove Theorem~\ref{thmMainResultLowerBoundDMC} as follows:
\begin{IEEEproof}[Proof of Theorem~\ref{thmMainResultLowerBoundDMC}]
Fix an $\varepsilon\in(0,1)$. Define
 \begin{equation}
a\triangleq \max\left\{\E_{p_{E_1}}[E_1^2],\max\left\{\left.(c(x))^2 e^{c(x)}\right| x\in \mathcal{X} \right\}\right\}. \label{defAinProofDMC}
\end{equation}
 Fix a sufficiently large~$n\ge 3$ such that \eqref{sufficientlyLargeNinProofDMC}, \eqref{sufficientlyLargeNinProof*DMC} and \eqref{sufficientlyLargeNinProof**DMC} hold.
 Define
\begin{equation}
m\triangleq \left\lceil \frac{6\sqrt{a n\log n}}{P} \right\rceil, \label{defmDMC}
\end{equation}
which specifies the number of time slots which are used for saving energy.
Consider the random code that uses the channel $m+n$ times as follows: \\
\textbf{Save-and-Transmit Random Codebook Construction}\\
Let $\mathbf{0}^m$ denote the length-$m$ zero tuple. Let $p_X$ be an arbitrary distribution on~$\mathcal{X}$ such that
 \begin{equation}
\E_{p_X}\left[c(X)\right]=P \label{defDistXDMC}.
 \end{equation}
 In addition, let $p_{X^n}$ be the distribution of the~$n$ independent copies of~$X$. Construct~$M$ i.i.d.\ random tuples denoted by $X^n(1), X^n(2), \ldots, X^n(M)$ such that $X^n(1)$ is distributed according to~$p_{X^n}$, where $M$ will be carefully chosen later when we evaluate the probability of decoding error. Define
 \begin{equation}
 \tilde X^{m+n}(i)\triangleq (\mathbf{0}^m, X^n(i)) \label{defTildeXmnDMC}
 \end{equation}
  for each $i\in\{1, 2, \ldots, M\}$ and construct the random codebook
\begin{equation}
\big\{ \tilde X^{m+n}(i) \,\big|\, i\in\{1, 2, \ldots, M\} \big\}. \label{defCodebookDMC}
\end{equation}
The codebook is revealed to both the encoder and the decoder.
Since the first~$m$ symbols of each random codeword $\tilde X^{m+n}(i)$ are zeros by \eqref{defTildeXmnDMC}, the source will just transmit~$0$ with probability one until time slot $m+1$ when the amount of energy $\sum_{k=1}^{m+1} E_k$ is available for encoding $\tilde X_{m+1}(W)\stackrel{\eqref{defTildeXmnDMC}}{=}X_1(W)$.
\\
\textbf{Encoding under the EH Constraints}\\
For each $i\in\{1, 2, \ldots, M\}$, let $\tilde X_k(i)$ be the $k^{\text{th}}$ element of $\tilde X^{m+n}(i)\stackrel{\eqref{defTildeXmnDMC}}{=} (\mathbf{0}^m, X^n(i))$ and construct recursively for $k=1, 2, \ldots, m+n$ the random variable
\begin{align}
\hat X_k(i, E^k)  \triangleq
\begin{cases}
\tilde X_k(i) &  \text{if $c(\tilde X_k(i)) \le \sum_{\ell=1}^k E_\ell - \sum_{\ell=1}^{k-1}c(\hat X_\ell(i, E^{\ell}))$,}\\
 0 & \text{otherwise.}
\end{cases} \label{defHatXkDMC}
\end{align}
To send message~$W$ which is uniformly distributed on $\{1, 2, \ldots, M\}$, the source transmits $\hat X_k(W, E^k)$ in time slot~$k$ for each $k\in\{1, 2, \ldots, m+n\}$.
Note that the source transmits~$0$ with probability one in the first~$m$ times slots by \eqref{defTildeXmnDMC} and \eqref{defHatXkDMC}, and the transmitted codeword $(\hat X_1(W, E^1), \hat X_2(W, E^2), \ldots, \hat X_{m+n}(W, E^{m+n}))$ satisfies the EH constraints \eqref{EHResultIntroDMC} by \eqref{defHatXkDMC}.
\\
\textbf{Threshold Decoding}\\
Upon receiving
$
\hat Y^{m+n} $ which is generated according to
\begin{align}
& p_{E^{m+n}, W, \hat X^{m+n}(W, E^{m+n}), \hat Y^{m+n}}(e^{m+n}, w, x^{m+n}, y^{m+n})\notag\\*
 & \stackrel{\eqref{memorylessStatementDM^*}}{=}p_{E^{m+n}, W, \hat X^{m+n}(W, E^{m+n})}(e^{m+n}, w, x^{m+n}) \prod_{k=1}^{m+n} q_{Y|X}(y_k|x_k) \label{defDistHatYmn}
\end{align}
where
\begin{align}
\hat X^{m+n}(W, E^{m+n})\triangleq (\hat X_1(W, E^1), \hat X_2(W, E^2), \ldots, \hat X_{m+n}(W, E^{m+n})) \label{defHatXmnDMC}
 \end{align}
 denotes the transmitted tuple specified in \eqref{defHatXkDMC},
 the destination constructs its subtuple denoted by $\bar Y^n$ by keeping only the last~$n$ symbols of $\hat Y^{m+n}$. Recalling that $q_{Y|X}$ denotes the channel law and $p_X$ was chosen to satisfy~\eqref{defDistXDMC}, we define
 the joint distribution
 \begin{equation}
p_{X,Y}\triangleq p_Xq_{Y|X}, \label{defJointDistributionDMC}
 \end{equation}
and define $p_{X^n, Y^n}$ to be the probability distribution of~$n$ independent copies of $(X,Y)$ distributed according to $p_{X,Y}$, i.e.,
 \begin{equation}
 p_{X^n, Y^n}(x^n, y^n) \triangleq \prod_{k=1}^n p_{X,Y}(x_k,y_k) \label{factorizationOfPXYDMC}
 \end{equation}
 for all $(x^n, y^n)\in\mathbb{R}^2$. Then, the decoder declares $\varphi(\bar Y^n) \in\{1, 2, \ldots, M\}$ (we write $\varphi(\bar Y^n)$ instead of  $\varphi(\hat Y^{m+n})$ to simplify notation) to be the transmitted message where $\varphi(\bar Y^{n})$ is the decoding function defined as follows:
 If there exists a unique index~$j$ such that
\begin{equation}
\log\left(\frac{p_{Y^n|X^n}(\bar Y^n| X^n(j))}{p_{Y^n}(\bar Y^n)}\right) > \log M + n^{\frac{1}{4}}, \label{decodingRuleDMC}
\end{equation}
then $\varphi(\bar Y^{n})$ is assigned the value~$j$.
Otherwise, $\varphi(\bar Y^{n})$ is assigned a random value uniformly distributed on $\{1,2, \ldots,M\}$.
\\
\textbf{Calculating the Probability of Violating  the EH Constraints}\\
Define $\bar X^n(W, E^{m+n})$ to be the tuple containing the last~$n$ symbols of $\hat X^{m+n}(W, E^{m+n})$.
Following similar procedures for proving \eqref{eqn6InCalculationErrorProb*}, we can obtain from \eqref{defHatXkDMC}, \eqref{defTildeXmnDMC}, Lemma~\ref{lemmaCharacteristicFunctionDMC}, \eqref{defAinProofDMC}, \eqref{sufficientlyLargeNinProofDMC}, \eqref{defmDMC} and \eqref{sufficientlyLargeNinProof**DMC} that
\begin{equation}
\Pr\left\{ \bar X^n(W, E^{m+n}) = X^{n}(W)\right\} \ge 1- \frac{\varepsilon}{2+\varepsilon}. \label{eqn6InCalculationErrorProbDMC}
\end{equation}
\textbf{Calculating the Probability of Decoding Error}\\
Choose $M$ to be the unique integer that satisfies
\begin{align}
\log (M+1)  &\ge n \E_{p_{X,Y}}\left[\log\left(\frac{p_{Y|X}(Y|X)}{p_{Y}(Y)}\right)\right] - \sqrt{\frac{(2+\varepsilon)n}{\varepsilon} \Var_{p_{X,Y}}\left[\log\left(\frac{p_{Y|X}(Y|X)}{p_{Y}(Y)}\right)\right]} - n^{\frac{1}{4}}\notag\\
&> \log M. \label{defChoiseOfMDMC}
\end{align}
Following similar proof steps from equation~\eqref{eqn6InCalculationErrorProb} to equation~\eqref{eqn9InCalculationErrorProb} for showing the existence of an $(n+m, M, \varepsilon)$-code for the AWGN EH channel that satisfies \eqref{eqn9InCalculationErrorProb}, 
we can show that the constructed code is an $(n+ m, M, \varepsilon)$-code that satisfies
\begin{align*}
\log (M+1) \ge n \E_{p_{X,Y}}\left[\log\left(\frac{p_{Y|X}(Y|X)}{p_{Y}(Y)}\right)\right]
 - \sqrt{\frac{(2+\varepsilon)n}{\varepsilon} \Var_{p_{X,Y}}\left[\log\left(\frac{p_{Y|X}(Y|X)}{p_{Y}(Y)}\right)\right]} - n^{\frac{1}{4}},
\end{align*}
which implies from \eqref{defJointDistributionDMC}, \eqref{defDistXDMC}, \eqref{defChannelInDefinition*DM} and \eqref{defVDMC} that
\begin{align}
\log (M+1) \ge n \E_{p_{X,Y}}\left[\log\left(\frac{p_{Y|X}(Y|X)}{p_{Y}(Y)}\right)\right]- \sqrt{\frac{(2+\varepsilon)n\mathrm{V}(P)}{\varepsilon}} - n^{\frac{1}{4}}, \label{eqn10InCalculationErrorProbDMC}
\end{align}
which then implies that
\begin{align}
1+\log M &\ge n \E_{p_{X,Y}}\left[\log\left(\frac{p_{Y|X}(Y|X)}{p_{Y}(Y)}\right)\right]- \sqrt{\frac{(2+\varepsilon)n\mathrm{V}(P)}{\varepsilon}} - n^{\frac{1}{4}}  \notag\\
& \ge \!((m\!+\!n)\!- \!m) \E_{p_{X,Y}}\!\! \left[  \log \! \left(\!\frac{p_{Y|X}(Y|X)}{p_{Y}(Y)}\!\right)\! \right]  - \sqrt{\frac{(2+\varepsilon)(m+n)\mathrm{V}(P)}{\varepsilon}} \! - \! (m+n)^{\frac{1}{4}} . \label{eqn11InCalculationErrorProbDMC}
\end{align}
Since
$
m \ge 0 $
and
$
 m \le \frac{6\sqrt{a n\log n}}{P} + 1$
by \eqref{defmDMC}, it follows that
\begin{align}
\frac{6\sqrt{a(m+n)\log(m+n)}}{P}+1 \ge \frac{6\sqrt{an\log n}}{P}+1 \ge m,
\end{align}
which implies from \eqref{eqn11InCalculationErrorProbDMC} and \eqref{defCDMC} that
\begin{align}
\log M
 & \ge (m+n)\E_{p_{X,Y}}\left[\log\left(\frac{p_{Y|X}(Y|X)}{p_{Y}(Y)}\right)\right] - \frac{6\mathrm{C}(P)\sqrt{a(m+n)\log(m+n)}}{P} \notag\\*
&\quad - \sqrt{\frac{(2+\varepsilon)(m+n)\mathrm{V}(P)}{\varepsilon}}\! -\! (m+n)^{\frac{1}{4}}\!-\mathrm{C}(P)\!-1 \, . \label{eqn12InCalculationErrorProbDMC}
\end{align}
The theorem then follows from \eqref{eqn12InCalculationErrorProbDMC}, \eqref{defAinProofDMC}, 
 \eqref{defmDMC}, \eqref{defDistXDMC} and \eqref{defCDMC} by letting $n^*\triangleq m+n$.
\end{IEEEproof}
 \section{Conclusion and Future Work} \label{sec:conclusion}
This paper has provided the first systematic study of finite blocklength achievable rates over noisy EH channels. We observe that the backoff from capacity at a finite blocklength $n$ is of the order $O( \sqrt{n^{-1} \, \log n} )$.
After the present work was submitted and placed on the arXiv, Shenoy and Sharma~\cite{ShenoySharma16} used Kolmogorov's inequality
(a consequence of Doob's maximal inequality)
 to bound the probability that the maximum (in time) of the difference between the harvested energy and the consumed energy (a martingale) exceeds a certain positive constant. They showed that the backoff term can be improved to $O(n^{-1/2}(\log n)^a)$ for any $a>0$. In fact, by combining our arguments and theirs, it is not difficult to further improve this backoff term to $O(n^{-1/2})$. This $O(n^{-1/2})$ scaling (rate of growth) in the second-order term is clearly optimal \cite{Strassen, PPV10}. At the same time, the development of new strategies for nailing down the constant is clearly a fertile avenue for future research. However, taking the $n$ EH constraints in~\eqref{EHResultIntro} into account to obtain a tighter bound from the meta-converse or its relaxed versions \cite[Sec.~III-E and III-F]{PPV10}
 (compared to the usual normal approximation)
 does not seem to be straightforward. 

\appendix
\section*{Proof of Lemma~\ref{lemmaCharacteristicFunction}}\label{appendixA}
We start with the following basic fact.
\begin{Proposition} \label{propositionTaylorExpansion}
For each non-negative real number~$x$,
\begin{equation}
1+x\le e^x \le 1+ x+ x^2 e^x/2 \label{propositionStatement1}
\end{equation}
and
\begin{equation}
1-x \le e^{-x} \le 1 - x+ x^2/2\,. \label{propositionStatement2}
\end{equation}
\end{Proposition}
\begin{IEEEproof}
For any real number $u\in \mathbb{R}$, it follows from Taylor's theorem that
\begin{equation}
e^u = 1 + u + u^2 e^{c}/2\label{taylorExpan}
\end{equation}
where $c$ is some number between $0$ and $u$. Inequalities \eqref{propositionStatement1} and \eqref{propositionStatement2} follow from \eqref{taylorExpan}.
\end{IEEEproof}
We are now ready to prove Lemma~\ref{lemmaCharacteristicFunction}.
\begin{IEEEproof}[Proof of Lemma~\ref{lemmaCharacteristicFunction}]
Fix a sufficiently large~$n$ that satisfies \eqref{lemmaCharFuncAssump3a} and fix a $k\in\{1, 2, \ldots, n\}$. To simplify notation, define
 \begin{equation}
\gamma_n\triangleq \frac{\log n}{a n} \label{defGammaN}
 \end{equation}
 where $a$ satisfies \eqref{lemmaCharFuncAssump2}.
 Consider the following chain of inequalities, where subscripts of probability and expectation terms are omitted for simplicity:
\begin{align}
\Pr\left\{ \sum_{\ell=1}^k X_\ell^2 \ge \sum_{\ell=1}^{m+k} E_\ell\right\}
& =   \Pr\left\{ e^{\sqrt{\gamma_n}\left(\sum_{\ell=1}^k X_\ell^2 - \sum_{\ell=1}^{m+k} E_\ell\right)} \ge 1 \right\} \notag\\
& \stackrel{\text{(a)}}{\le} \E\left[ e^{\sqrt{\gamma_n}\left(\sum_{\ell=1}^k X_\ell^2 - \sum_{\ell=1}^{m+k} E_\ell\right)}  \right]\notag \\
& \stackrel{\text{(b)}}{=} \left(\E\left[e^{\sqrt{\gamma_n}X_1^2}\right]\right)^k \left(\E\left[e^{-\sqrt{\gamma_n}E_1}\right]\right)^{m+k} \notag\\
& \stackrel{\text{(c)}}{\le} \left( 1+\sqrt{\gamma_n}\E\left[X_1^2\right] + \frac{\gamma_n}{2}\E\left[X_1^4 e^{\sqrt{\gamma_n}X_1^2}\right] \right)^k \left( 1-\sqrt{\gamma_n}\E\left[E_1\right] + \frac{\gamma_n}{2}\E[E_1^2] \right)^{m+k} \notag\\
&\stackrel{\text{(d)}}{\le} \left( 1+\sqrt{\gamma_n}\E\left[X_1^2\right] + \frac{\gamma_n}{2} \E\left[X_1^4 e^{\lambda X_1^2}\right] \right)^k \left(1-\sqrt{\gamma_n}\E\left[E_1\right] + \frac{\gamma_n}{2}\E[E_1^2] \right)^{m+k} \notag\\
& \stackrel{\text{(e)}}{\le}  \left( 1+\sqrt{\gamma_n}P + \frac{ a\gamma_n}{2} \right)^k \left( 1-\sqrt{\gamma_n}P + \frac{ a\gamma_n}{2} \right)^{m+k}\notag \\
&\stackrel{\text{(f)}}{\le}  \left(e^{\sqrt{\gamma_n}P + \frac{ a\gamma_n}{2}}\right)^k \left( e^{-\sqrt{\gamma_n}P + \frac{ a\gamma_n}{2}} \right)^{m+k}\notag \\
& =  e^{ka\gamma_n}  \left( e^{-\sqrt{\gamma_n}P + \frac{a\gamma_n}{2}} \right)^{m} \notag\\
&\stackrel{\text{(g)}}{\le}  e^{ka\gamma_n-\frac{mP\sqrt{\gamma_n}}{2}}, \label{eqnInLemmaCharFunc}
\end{align}
where
\begin{enumerate}
\item[(a)] follows from Markov's inequality.
\item[(b)] follows from the facts that $\{X_k\}_{k=1}^n$ and $\{E_k\}_{k=1}^{m+n}$ are two sequences of i.i.d\ random variables and $X^n$ and $E^{m+n}$ are independent.
    \item[(c)] follows from the upper bounds in Proposition~\ref{propositionTaylorExpansion}.
    \item[(d)] uses the following fact due to \eqref{defGammaN} and \eqref{lemmaCharFuncAssump3a}: $\sqrt{\gamma_n}\le \lambda$.
    \item[(e)] follows from \eqref{lemmaCharFuncAssump1} and \eqref{lemmaCharFuncAssump2}.
     \item[(f)] follows from the lower bounds in Proposition~\ref{propositionTaylorExpansion}.
      \item[(g)] follows from the fact that
    \begin{equation}
\!\!\!\!\!\!  \frac{a\gamma_n}{2} \!=\! \frac{a\sqrt{\gamma_n}}{P}\cdot\frac{P\sqrt{\gamma_n}}{2} \stackrel{\eqref{defGammaN}}{=} \!\sqrt{\frac{a \log n}{n P^2}}\cdot\frac{P\sqrt{\gamma_n}}{2} \stackrel{\eqref{lemmaCharFuncAssump3a}}{\le} \!  \frac{P\sqrt{\gamma_n}}{2}\, .
    \end{equation}
\end{enumerate}
Consequently, it follows from the union bound that
\begin{align}
\Pr\left\{\bigcup_{k=1}^n \left\{ \sum_{\ell=1}^k X_\ell^2 \ge \sum_{\ell=1}^{m+k} E_\ell\right\}\right\}
& \le \sum_{k=1}^n \Pr\left\{ \sum_{\ell=1}^k X_\ell^2 \ge \sum_{\ell=1}^{m+k} E_\ell\right\}\notag\\
& \stackrel{\eqref{eqnInLemmaCharFunc}}{\le} \sum_{k=1}^n e^{ka\gamma_n-\frac{mP}{2}\sqrt{\gamma_n}} \notag\\
& = e^{-\frac{mP}{2}\sqrt{\gamma_n}} \left(\frac{e^{a\gamma_n}(e^{an\gamma_n}-1)}{e^{a\gamma_n}-1}\right) \notag\\
& \stackrel{\eqref{propositionStatement1}}{\le}e^{-\frac{mP}{2}\sqrt{\gamma_n}} \left(\frac{e^{a\gamma_n}(e^{a n\gamma_n}-1)}{a\gamma_n}\right) \notag\\
& <  e^{-\frac{mP}{2}\sqrt{\gamma_n}} \left(\frac{e^{a\gamma_n}(e^{a n\gamma_n})}{a\gamma_n}\right) \notag\\
& \stackrel{\text{(a)}}{\le} e^{-\frac{mP}{2}\sqrt{\gamma_n}} \left(\frac{e^{0.4}(e^{a n\gamma_n})}{a\gamma_n}\right) \notag\\
&  \stackrel{\eqref{defGammaN}}{=}   \left(\frac{e^{0.4}}{\log n}\right)e^{2\log n-\frac{mP}{2}\sqrt{\frac{\log n}{a n}}}
\end{align}
where (a) follows from the fact that for $n\ge 3$,
$
a\gamma_n\stackrel{\eqref{defGammaN}}{=}\frac{\log n}{n} \le \frac{\log 3}{3} < 0.4 .$
\end{IEEEproof}

\section*{Acknowledgements}
Silas Fong and Vincent Tan gratefully acknowledge financial support from  the National University of Singapore (NUS) under   grant R-263-000-A98-750/133 and  NUS Young Investigator Award   R-263-000-B37-133.
Jing Yang is supported by United States National Science Foundation (NSF) under grants ECCS-1405403 and ECCS-1454471.
%

%
%
%
%
%




\end{document}